\documentclass[aps,twocolumn,superscriptaddress]{revtex4}
\usepackage{amsmath}
\usepackage{amssymb}
\usepackage{graphicx}
\usepackage{verbatim}
\usepackage{color}
\usepackage{bbm}           

\usepackage[
	colorlinks=true,
	citecolor=black,
	linkcolor=black,
	urlcolor=black,
	hypertexnames=false]{hyperref}

\renewcommand{\eqref}[1]{Eq.~(\ref{#1})}
\newcommand{\ket}[1]{\left|#1\right\rangle}    % <#1|
\newcommand{\bra}[1]{\left\langle#1\right|}    % |#1>

\DeclareMathOperator\sinc{sinc}

\begin{document}

%\preprint{}

%\title{Vector Space Integration for Dark Matter Scattering}
%\title{Rescuing Anisotropic Direct Detection With Harmonics and Factorization}
\title{Partial Rate Matrix for Dark Matter Scattering}

\author{Benjamin Lillard}
\email[]{blillard@uoregon.edu}
\affiliation{Institute for Fundamental Science and Department of Physics,\\ Willamette Hall, University of Oregon, Eugene, OR 97401, U.S.A.}

\date{\today}

\begin{abstract}
I present a highly efficient integration method for scattering calculations,
and  a ``partial rate matrix'' that encodes the scattering rate as a function of the $SO(3)$ orientation of the detector.
This replaces the original multidimensional rate integral with a simple exercise in vector multiplication, speeding up the rate calculation by a factor of around $10^8$.
I include a scheme to fully factorize the dark matter particle model, its astrophysical velocity distribution, and the properties of the target material from each other, enabling efficient calculation of the partial rate matrix even in studies comparing large sets of these input functions. 
This is now the only sensible way to evaluate the dark matter scattering rate in anisotropic detector materials.
It is straightforward to generalize this method to other difficult but linear problems.

\end{abstract}

%\keywords{}

\maketitle

%%%%%%%%%%%%%%%%%%%%%%%%%%%%%%%%%%%%%%%%%%%%%%%%%%%%%%%%%%%%%%%%%%%%%%%%%%%%%%%%%%%%
%%%%%%%%%%%%%%%%%%%%%%%%%%%%%%%%%%%%%%%%%%%%%%%%%%%%%%%%%%%%%%%%%%%%%%%%%%%%%%%%%%%%
\section{\label{sec:intro} Introduction}

Dark matter direct detection has become increasingly complicated. The search for low mass (e.g. sub-GeV) dark matter (DM) is driving a search for detector materials with lower energy thresholds, including 
semiconductors~\cite{SuperCDMS:2018mne,DAMIC:2019dcn,EDELWEISS:2020fxc,SENSEI:2020dpa,DAMIC-M:2023gxo}, 
molecular targets~\cite{Blanco:2019lrf,Blanco:2021hlm,Blanco:2022pkt}, 
and a variety of anisotropic materials~\cite{Hochberg:2016ntt,Budnik:2017sbu,Hochberg:2017wce,Coskuner:2019odd,Geilhufe:2019ndy,Griffin:2020lgd,Coskuner:2021qxo,Boyd:2022tcn,Catena:2023qkj,Catena:2023awl}.
Predicting the DM scattering rate in such experiments requires input from condensed matter or physical chemistry, complicating what for liquid noble elements~\cite{XENON:2018voc,PandaX-II:2021nsg,LZ:2022ufs,XENON:2023sxq,DarkSide:2018ppu,XENON:2022ltv} would have been a relatively simple calculation~\cite{Lewin:1995rx}.

Materials with directionally dependent scattering rates have the capability to distinguish the DM signal from any Standard Model (SM) backgrounds, but at a computational cost: the scattering rate now depends on integrals over the three-dimensional momentum transfer and DM velocity vectors, $\vec{q}$ and $\vec{v}$, rather than simply  $|\vec{q}|$ and $|\vec{v}|$. 
Furthermore, the rate depends on the detector orientation, requiring a scan over elements of the $SO(3)$ rotation group. So, not only is the rate integral more numerically challenging, it must also be repeated very many times, e.g.~to identify the optimal detector orientations to maximize (or minimize) the modulation amplitude.
To capture the time dependence of the velocity distribution, e.g.~the annual variation~\cite{Drukier:1986tm,Freese:1987wu, Lee:2015qva} generated by the Earth's revolution around the Sun~\cite{10.1093/mnras/4.19.168}, the rate calculation must also be repeated for many different lab-frame velocity distributions.

There are large uncertainties on the astrophysical dark matter velocity distribution, which probably contains anisotropic remnants (e.g.~unvirialized streams) from the more recent galactic mergers~\cite{Diemand:2008in,Klypin:2010qw,Guedes:2011ux,Hopkins:2017ycn,Kuhlen:2012fz,Necib:2018igl,Riley:2018lbh}, in some cases altering the direct detection constraints~\cite{Wu:2019nhd,Radick:2020qip,Maity:2020wic,Buckley:2022tjy,Maity:2022enp}.
Despite the fact that the detector response is governed by SM physics, these SM form factors are often subject to their own uncertainties~\cite{Knapen:2021run,Knapen:2021bwg,Griffin:2021znd}.
Due in part to the computational difficulty, these uncertainties are rarely propagated through the direct detection rate calculation. 

A complete analysis requires simultaneous scans over $N_{\mathcal R}$ detector orientations; $N_{g\chi}$ different lab-frame velocity distributions; 
$N_\text{DM}$ particle models of the DM, specifying the DM mass $m_\chi$ and scattering form factor $F_\text{DM}$; and $N_{f_S}$ different models for the detector physics (e.g.~for different target materials or final states). 
In the standard approach, the rate integral must be evaluated $N_{\mathcal R}  N_{g\chi}  N_\text{DM}  N_{f_S}$ many times, which soon becomes prohibitively expensive~\footnote{In ref.~\cite{Blanco:2021hlm}, for example, our computational budget limited us to $N_{\mathcal R} = 48$, in a standard analysis with $N_{g_\chi} = 1$.}.

In this paper I present a method of ``vector space integration'' that factorizes the astrophysics, the DM particle physics, and the SM physics of the target material by projecting the velocity distribution $g_\chi(\vec v)$ and the SM momentum form factor $f_S^2(\vec q)$ onto bases of orthogonal functions. After an initial, numerically intensive stage where each object is projected onto its respective vector space, the scattering rate is given by simple matrix multiplication on the vectors $\ket{g_\chi}$ and $\ket{f_S^2}$.
I use spherical harmonics as part of the basis, so that detector rotations are encoded by the action of the Wigner $D$ matrix on the basis functions.

This solution to the $N^4$ scaling problem is shockingly effective~\footnote{Note that this $N^4$ scaling problem is of a completely different nature than the $N^4$ scaling problem encountered in computational chemistry in the electron exchange correlation functions. The resolution of the identity technique, which ameliorates the $N^4$ scaling in computational chemistry, does not offer a solution to the $N_{\mathcal R}  N_{g\chi}  N_\text{DM}  N_{f_S}$ scaling of the direct detection rate calculations. See the Supplemental Material for a more detailed discussion.}. The vector space method requires only $N_{g\chi} + N_{f_S}$ repetitions of the harder, numerical integration part of the calculation, leading to a substantial reduction in the calculation time if any two or more of the various $N_i$ are large. 
Even in simple isotropic materials ($N_\mathcal{R} = 1$), the factorization of $N_{g\chi}$, $N_{f_S}$ and $N_\text{DM}$ can reduce the computational difficulty by multiple orders of magnitude. 
For anisotropic detectors in the limit of large $N_\mathcal{R}$, Ref.~\cite{Lillard:2023cyy} finds that the computation time is reduced by a factor of approximately $10^8$.

Whether the impact of this technology is greatest for direct detection or for some other scientific field remains to be seen. 
The vector space integration method is easily generalized to other contexts, where the integrand depends linearly on any number of input functions (in this case $g_\chi$ and $f_S^2$), along with some operator that depends on the coordinates and any external parameters (e.g.~$\vec v$, $\vec q$ and $m_\chi$). 
If any symmetries of the integrand can be exploited, i.e.~by choosing basis functions that transform as representations of that symmetry, then the vector space integration can provide the same extreme reduction in computation time that is demonstrated here for direct detection.

\section{Integration In Vector Space} \label{sec:method}

The spin-independent DM scattering rate in a target containing $N_T$ SM particles is~\cite{Essig:2015cda,Trickle:2019nya,Hochberg:2021pkt}:
\begin{align}
R &= N_T \frac{\rho_\chi}{m_\chi} \int\! dE\, d^3 q\,d^3 v\, g_\chi(\vec v)  \, V_\text{cell} S(\vec q, E)  
\nonumber\\&~~~ \times \delta\! \left( E + \frac{q^2}{2m_\chi} - \vec q \cdot \vec v \right) 
\left( \frac{\bar\sigma_0 F_\text{DM}^2(q) }{8\pi^2 \mu_\chi^2} \right) ,
\label{eq:rateE}
\end{align}
where $\vec q$ and $E$ are the momentum and energy transferred from DM to the SM particle; $g_\chi(\vec v)$ is a nonrelativistic DM velocity distribution, normalized to $\int\! d^3 v \, g_\chi(\vec v) = 1$; $\rho_\chi$ is the local DM mass density; $\mu_\chi = m_\chi m_\text{SM} / (m_\chi + m_\text{SM})$ is the reduced mass of the SM--DM system, for SM particle mass $m_\text{SM}$; and $V_\text{cell} S(\vec q, \omega)$ is the dynamic structure function~\cite{Trickle:2019nya,Hochberg:2021pkt}
which encodes the properties of the target material, with volume normalization factor $V_\text{cell}$.
The cross section is factored into a constant $\bar\sigma_0$ and a momentum-dependent function $F^2_\text{DM}(q)$, which is normalized to $F_\text{DM}(q_0) \equiv 1$ at some reference momentum $q_0$, such that $\sigma(\vec q)= \sigma_0 F_\text{DM}^2(\vec q)$.

\eqref{eq:rateE} is appropriate for scattering in semiconductors, nuclear recoil, and atomic ionization, with a continuum of final states labeled by the energy transfer $E$.
In other scenarios, e.g.~molecular scattering to bound excited states~\cite{Blanco:2019lrf,Blanco:2021hlm,Blanco:2022pkt}, the final states $s$ take discrete energies $\Delta E_s$, and $V_\text{cell} S$ is replaced by $2 \pi f_s^2(\vec q) \, \delta(E - \Delta E_s) $:
\begin{align}
R &=N_T  \rho_\chi \bar\sigma_0  \sum_s  \int\!  d^3 q\,d^3 v\, g_\chi(\vec v)  \, f_s^2(\vec q)
\nonumber\\&\hspace{2em} \times   \frac{  F_\text{DM}^2(q) }{4\pi \mu_\chi^2 m_\chi}  \delta\! \left( \Delta E_s + \frac{q^2}{2m_\chi} - \vec q \cdot \vec v \right) 
.
\label{eq:rate}
\end{align}
See e.g.~\cite{Essig:2011nj,Essig:2015cda,Trickle:2019nya,Blanco:2021hlm,Hochberg:2021pkt} for details.
In the following, I write both $ S(\vec q, E) V_\text{cell}/2\pi$ and $f_s^2(\vec q) \delta(E - \Delta E)$ as $f_S^2(\vec q, E)$, a material-specific form factor with units of inverse energy.

I project the functions of $\vec v$ and $(\vec q, E)$ onto vector spaces spanned by complete and orthogonal basis functions $\ket{\phi}$ and $\ket{\varphi}$:
\begin{align}
\ket{\phi_{n \ell m} (\vec v)} &= r_n^{(v)}(v) Y_{\ell m}(\hat v) , 
\label{basis:phi}
\\
\ket{\varphi_{j n \ell m} (\vec q, E) } &= r_{jn}^{(q)}(q, E) Y_{\ell m }(\hat q),
\label{basis:varphi}
\end{align}
where $Y_{\ell m}(\hat n)$ are the real spherical harmonics, 
\begin{align}  
Y_{\ell m}(\Omega) &\equiv
\left\{ \begin{array}{l c c}
\sqrt{2} \, \text{Im}\, Y_{\ell}^{|m|}(\Omega)  
&& \text{for } m < 0,
\\[\medskipamount]
Y_\ell^0(\Omega)  && \text{for } m = 0,
\\[\medskipamount]
\sqrt{2} \, \text{Re}\,Y_\ell^m (\Omega) 
&& \text{for } m > 0 ,
\end{array}  
\right.
\label{Ylm:realdef}
\end{align}
and $r_n(v)$ and $r_{jn}(q, E)$ are real-valued orthogonal functions.
Here I leave the form of the radial basis functions unspecified. A companion paper~\cite{Lillard:2023cyy} introduces a basis of spherical wavelets specifically designed for this application.

Using a bra-ket notation for the basis vectors, the vector space representations of $g_\chi$ and $f_S$ are:
\begin{align}
g_\chi &= \sum_{\phi} \langle g_\chi | \phi \rangle \bra{\phi},
%\\
&
f_S^2 &= \sum_{\varphi} \left\langle \varphi \big| f_S^2 \right\rangle \ket{\varphi}.
\label{fdet:vector}
\end{align}
Inner products take the form 
\begin{align}
\langle f | \phi \rangle & = \int\! d^3v\, f(\vec v) \phi(\vec v),
\\
\langle  f| \varphi \rangle &= \int\! d E\, d^3q \, f(\vec q, E) \, \varphi(\vec q, E).
\label{eq:inner}
\end{align}
For $L^2$ normalized basis functions, $\langle \phi | \phi \rangle = \langle \varphi | \varphi \rangle =1$, $\ket{g_\chi}$ and $\ket{f_S^2}$ are Hilbert space representations of the original functions.
The DM--SM scattering is represented by a matrix $\mathcal M$, with coefficients:
\begin{align}
\mathcal{M}_\phi^\varphi &\equiv 
\left\langle \! \phi (\vec v) \left| \frac{F_\text{DM}^2 (q) }{4\pi m_\chi \mu_\chi^2} \delta\!\left(E + \frac{q^2}{2 m_\chi } - \vec q \cdot \vec v \right) \right| \varphi (\vec q, E) \! \right\rangle ,
\label{mathM:generic}
\end{align}
so that the scattering rate $R$ is given by
\begin{align}
R(\mathcal R) = N_T \rho_\chi \bar\sigma_0 \sum_{\phi} \sum_{\varphi} \langle g_\chi | \phi \rangle \cdot \mathcal M_{\phi}^\varphi \cdot \langle \varphi | f_S^2(\mathcal R) \rangle .
\label{rate:mathM}
\end{align}
For anisotropic materials, the rate $R$ depends on the orientation of the detector, as indicated above by writing $R$ and $f_S^2$ as functions of $\mathcal R \in SO(3)$.
Later in this Section, 
I show that $R(\mathcal R)$ can be evaluated extremely quickly once put into the form:
\begin{align}
 R(\mathcal R) &= N_T \rho_\chi \bar\sigma_0 \sum_{\ell, m, m'}
 \langle \ell m | \mathcal R| \ell m' \rangle  K^{(\ell)}_{m m'} \, ,
\end{align}
with $-\ell \leq m, m' \leq \ell$. 
Here $K^{(\ell)}_{m m'}[g_\chi, f_S^2]$ is the \emph{partial rate matrix}, defined as:
\begin{align}
K^{(\ell)}_{m m'} &\equiv \int\!  \frac{dE\,qdq\, vdv}{2 \mu_\chi^2 m_\chi} P_\ell\!\left(\frac{v_\text{min} }{v}\right) g_{\ell m} f_{\ell m'}^2  F_\text{DM}^2   ,
\end{align}
where $g_{\ell m}(|\vec v|) = \langle \ell m | g_\chi \rangle$ and $f_{\ell m'}^2(|\vec q|, E) = \langle \ell m' | f_S^2 \rangle$ are the projections of $g_\chi(\vec v)$ and $f_S^2(\vec q, E)$ onto spherical harmonic modes, 
and $v_\text{min}(q, E) = E/q + q/(2 m_\chi)$.

\subsection{Spherical Harmonics and Kinematics} \label{sec:kinematics}

At first glance, $\mathcal M_\phi^\varphi$ looks like the most numerically expensive object in \eqref{rate:mathM}: it must be integrated over $\vec q$ and $\vec v$ (and possibly $E$), for every pair of $(\phi, \varphi)$ basis functions. 
Fortunately, aside from $F_\text{DM}$ and the energy-conserving $\delta$ function, 
the inputs to $\mathcal M$ are completely determined by the choice of basis functions.
$\mathcal M_\phi^\varphi$ simplifies dramatically in the spherical harmonic basis, especially when $F_\text{DM}(\vec q)$ is isotropic (e.g.~for spin-independent or spin-averaged cross sections).
With $g_\chi(\vec v)$ provided in the lab frame, and with $F_\text{DM}(\vec q) = F_\text{DM}(q)$, the operator in \eqref{mathM:generic} is rotationally invariant. Consequently, $\mathcal M$ is diagonal in the angular indices $\ell$ and $m$.
This is shown explicitly in Ref.~\cite{Lillard:2023cyy} using a completeness relation of the Legendre polynomials and real spherical harmonics. 

Returning the $(n \ell m)$ and $(j n' \ell' m')$ indices to $\ket{\phi}$ and $\ket{\varphi}$, \eqref{mathM:generic}  simplifies to~\footnote{This simplification $\mathcal M \rightarrow  I^{(\ell)}$ is a special case of the Wigner--Eckhart theorem, applied to the rank-0 spherical operator in \eqref{mathM:generic}. }:
\begin{align}
\mathcal M^{j n' \ell' m'}_{n \ell m}  
&\equiv \delta_\ell^{\ell'} \delta_m^{m'} \cdot \,  I^{(\ell)}_{n, j n'}(m_\chi, F_\text{DM}) ,
\label{eq:magic}
\\
 I^{(\ell)}_{n, j n'} & = \int_0^\infty \! dE\,qdq \frac{F_\text{DM}^2(q) }{2 m_\chi \mu^2} r^{(q)}_{j n'}(q, E)
\nonumber\\&~~~ \times  \int_{v_\text{min}(q, E) }^\infty \! vdv\, P_\ell\!\left( \frac{v_\text{min} }{v} \right) r^{(v)}_n(v), 
\label{eq:mathcalI}
\end{align}
where $P_\ell$ is the $\ell$th Legendre polynomial, and $v_\text{min}(q, E)$ is defined in the usual way, 
\begin{align}
v_\text{min}(q, E) \equiv \frac{E}{q} + \frac{q}{2 m_\chi} .
\label{eq:vmin}
\end{align}
Note that while the integrand of $I^{(\ell)}$ may diverge as $q\rightarrow 0$, e.g.~if $F_\text{DM} \propto 1/q^2$, a nonzero energy threshold $E \geq \Delta E$ ensures that $q=0$ is outside the integration region for any finite $v$, so that $I^{(\ell)}$ remains finite~\footnote{In principle, the radial basis functions $r^{(v)}_n(v)$ can extend to $+\infty$, as long as they vanish quickly enough to cancel any $1/q^n$ poles in the momentum integral, e.g.~as $r_n \propto e^{-v^2/v_0^2}$. Recall that the scattering rate \eqref{eq:rateE} assumes the nonrelativistic limit, $v \ll c$, so any support at $v \sim \mathcal O(0.1\,c)$ should at least be vanishingly small. Likewise, the physical $g_\chi$ is not expected to have any support at $v > c$, for more fundamental reasons.}.  

For an isotropic detector, only the $\ell = 0$ terms in \eqref{eq:magic} contribute to the scattering rate.

\subsection{Rotations}

Next, consider rotations of the detector, $f_S^2(\mathcal R)$ for $\mathcal R \in SO(3)$. Spherical harmonics of fixed $\ell$ form a $(2\ell + 1)$ dimensional representation of the rotation group, so the action of the rotation operator $\mathcal R$ on the basis functions $\ket{j n \ell m}$ is given by the Wigner $D$-matrix~\cite{wignerD}. Defining a $G^{(\ell)}$ matrix analogous to the Wigner $D^{(\ell)}$ for real spherical harmonics $\ket{\ell m} = Y_{\ell m}$,
\begin{align}
G^{(\ell)}_{m' m}(\mathcal R) \equiv \langle \ell m' | \mathcal R | \ell m \rangle, 
\label{eq:WignerG}
\end{align}
the scattering rate for a rotated detector is given by
\begin{align}
R(\mathcal R) & = N_T \rho_\chi \bar\sigma_0 \sum_{\ell} \sum_{n, n' j} \sum_{m, m'} \langle g_\chi | n \ell m \rangle 
\nonumber\\&~~~~ \times   I^{(\ell)}_{n, n' j} \cdot G^{(\ell)}_{m m'}(\mathcal R) \cdot \langle j n' \ell m' | f_S^2 \rangle .
\label{rate:mathI}
\end{align}
Although \eqref{eq:WignerG} is written in the form of an angular integral, its coefficients can be found directly from any explicit representation of the complex Wigner $D^{(\ell)}$ matrix.
No integration is necessary.

\subsection{Partial Rate Matrix} \label{sec:partialrate}

As the penultimate step in the calculation,
\eqref{rate:mathI} is organized into a series of partial rate matrices $K^{(\ell)}$ by completing the sums over $n, n'$ and $j$:
\begin{align}
 K^{(\ell)}_{m , m'}  &= \sum_{j, n, n'} \langle g_\chi | n \ell m \rangle \,  I^{(j, \ell)}_{n, n' } \, \langle j n' \ell m' | f_S^{2} \rangle ,
 \label{eq:mcalK} 
 \\
 R(\mathcal R) &= N_T \rho_\chi \bar\sigma_0 \sum_{\ell = 0}^{\infty} \sum_{m = -\ell}^{\ell}  \sum_{m' = -\ell}^{\ell}  G^{(\ell)}_{m m'}(\mathcal R)  K^{(\ell)}_{m m'} .
\label{eq:rateTrace}
\end{align}
Each combination of $g_\chi$, $f_S^2$, $m_\chi$ and $F_\text{DM}$ generates different $K^{(\ell)}$, each of which contains all of the information that might be extracted by rotating the detector and measuring the rate.

The angular-diagonal structure of \eqref{eq:mathcalI} separates the rate into contributions from fixed $\ell$, $R = \sum_\ell R^{(\ell)}$. States of different radial index $n,n'$ are mixed by $I^{(\ell)}$, and the rotation matrices $G^{(\ell)}$ mix the $m,m'$ of fixed $\ell$; but the different $\ell$ modes do not mix. 
The residual diagonalization in $\ell$  is highly relevant to the detector design: for example, to resolve a feature in $g_\chi$ at a specific $\ell$, the detector form factor must have support at that same $\ell$. 
Likewise, two detector materials can provide complementary information if their lowest $\ell$ form factors are substantially different. 
This feature also has a practical benefit: how the partial rate $R^{(\ell)}$ falls as a function of $\ell$ will indicate when the expansion in $\ell$ can be terminated.

\section{Convergence and Timing}

If the number of rotations in the analysis $N_\mathcal{R}$ is large, 
the time spent calculating the partial rate matrix $K^{(\ell)}_{m m'}$ quickly pays off. 
 \eqref{eq:rateTrace} can be evaluated in a few microseconds as the dot product between two vectors $G$ and $K$, but evaluating \eqref{eq:rateE} or \eqref{eq:rate} at similar precision for the same simple models can take several minutes.
This is the origin of the factor of $10^8$ cited from Ref.~\cite{Lillard:2023cyy}.
For realistic detector models with numerically-defined $f_S^2$, the ratio in evaluation times may be even larger.

\begin{figure}
\centering
\includegraphics[width=0.48\textwidth]{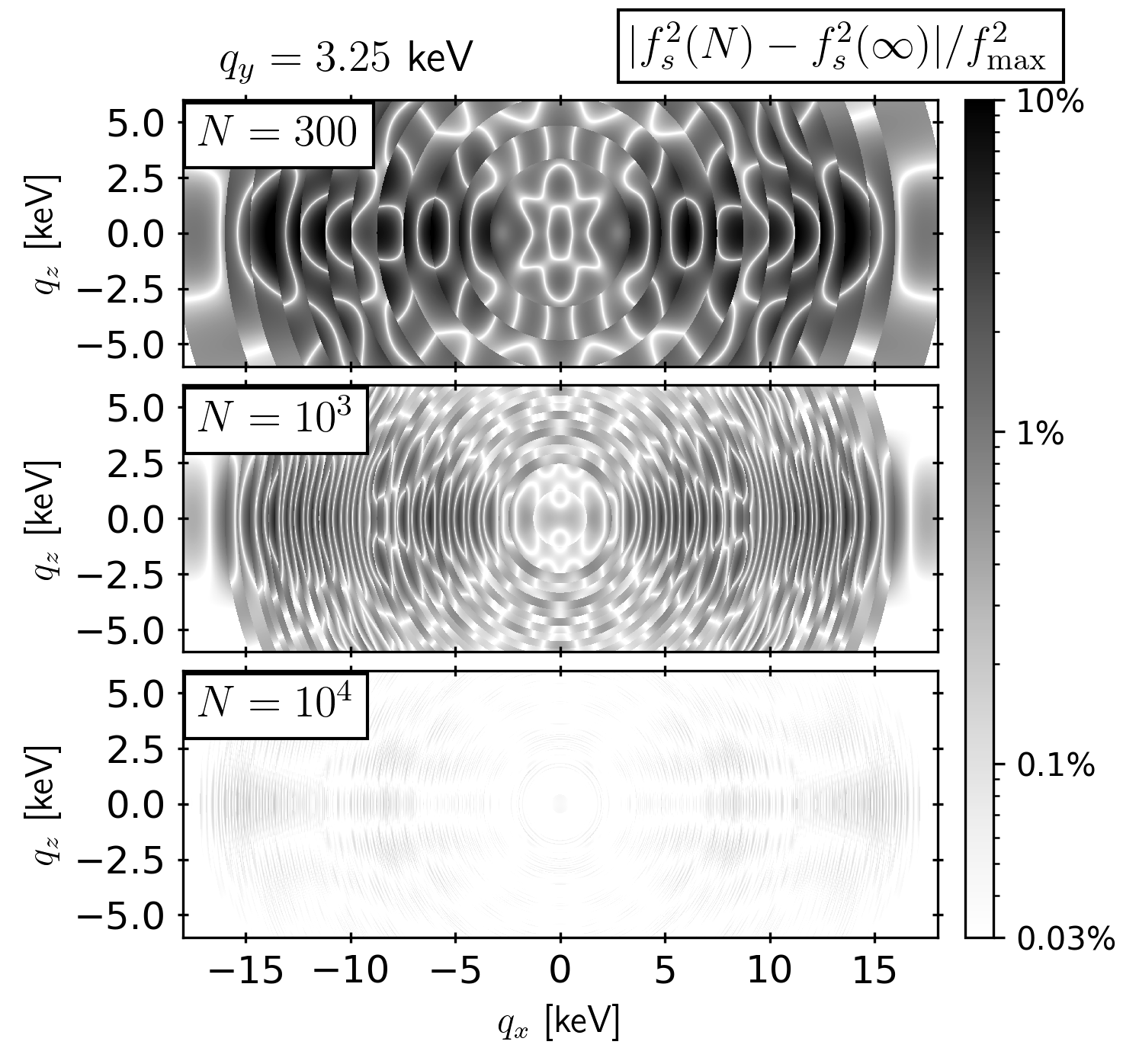}
\caption{
Demonstrating the convergence of a basis function expansion $f_s^2 \simeq \sum_\varphi \langle \varphi | f^2_s \rangle \ket{\varphi}$, summed over the $N = 300, 10^3, 10^4$ largest coefficients. 
Here $\ket{\varphi}$ are the spherical wavelets introduced in~\cite{Lillard:2023cyy}.
This example is for a particle of mass $m_e$ in a rectangular box with sides $(L_x, L_y, L_z) = (4, 7, 10)\times (\alpha m_e)^{-1}$, for the transition to one of the higher excited states. 
Each plot shows the absolute error in $f_s^2(\vec q)$, normalized by $f_\text{max}^2 \equiv \text{max}(f_s^2)$, for momenta on the plane $q_y = 3.25\, \text{keV}$ (which coincides with the maximum).
}
\label{fig:accuracy}
\end{figure}

First, it should be shown that the sums over $n, n'$ and $\ell,m,m'$ can be truncated, without an unacceptable loss of precision.
As a brief demonstration,  
Fig.~\ref{fig:accuracy} shows three versions of $\ket{f_S^2}$ using only the $N_q = 300, 10^3, 10^4$ largest coefficients (respectively), selected from a much larger initial set with $\ell_\text{max} = 36$ and $N_n^{(q)} = 2^{10}$. The example $f_S^2(\vec q)$ is for a highly excited state in a particle-in-a-box model taken from Ref.~\cite{Lillard:2023cyy}. (See  Appendix~\ref{appx:B} for further detail.) 
Each panel in Fig.~\ref{fig:accuracy} shows the difference between $f_S^2(\vec q)$ and $\sum_{\varphi} \langle \varphi | f_S^2 \rangle \varphi(\vec q)$, for a 2d slice in momentum space. 
With $N=300$ the local error exceeds 10\% in some places, but $N = 10^4$ is sufficient to reproduce the original function within $\sim 0.1\%$, 
which is negligible compared to e.g.~the uncertainty on $g_\chi(\vec v)$.
This is qualitative proof that the basis expansion converges sufficiently rapidly.
Ref.~\cite{Lillard:2023cyy} arrives at the same conclusion, after a more comprehensive study.

Including the intermediate steps, the total amount of time needed to complete the rate calculation is now:
\begin{align}
T_\text{tot} &= N_{g_\chi} N_v T_\text{proj.}^{(v)} +  N_{f_S}  N_q T_\text{proj.}^{(q)} + N_{I} T_{I} 
\nonumber\\&~~
+ N_{g_\chi} N_{f_S} N_\text{DM} \left( T_{ K}  + N_\mathcal{R} T_\text{Tr} \right) ,
\label{time:tot}
\end{align}
where $T_\text{proj,}^{(q,v)}$ is the average integration time for a single $\langle \varphi | f_S^2 \rangle$ or $\langle g_\chi | \phi\rangle$ inner product; $T_{I}$ is the evaluation time for one element of \eqref{eq:mathcalI};
$T_K$ is the time needed to assemble the partial rate matrix in \eqref{eq:mcalK} for one combination of $g_\chi$, $f_S^2$ and DM particle model; and $T_\text{Tr}$ is the evaluation time for the $\ell m m'$ sum in \eqref{eq:rateTrace}. 
From Ref.~\cite{Lillard:2023cyy}, typical values for $T_i$ are~\footnote{This $T_\text{proj.}$ is the integration time for $\langle \varphi| f_s^2\rangle$. In Ref.~\cite{Lillard:2023cyy}, the example velocity distribution is a sum of gaussian functions, which allows the angular integrals $\langle g_\chi | \phi \rangle$ to be performed analytically. As a result, $T_\text{proj.}^{(v)} < 0.01$\,s for this example. This favorable result is not quoted in the main text because it is not generic: it only applies to functions $g_\chi$ or $f_S^2$ that are given as the sum of individual spherical gaussians (centered at any points in $\vec v$ or $\vec q$).}:
\begin{align}
T_\text{proj.} &\sim \text{few seconds},
&
T_{I} &\sim 10^{-3} \, \text{s} ,
\\
T_{K} &\sim \ell_\text{max}^3 \cdot 10^{-5} \, \text{s}, 
&
T_\text{Tr} &\lesssim  \ell_\text{max}^3 \cdot 10^{-9}  \,\text{s} . 
\label{eq:forGPU}
\end{align}
In terms of the number of radial modes $N_n^{(q,v)}$, final energy states or energy basis functions $N_j$, and the number of values of $\ell$, $N_\ell$,
\begin{align}
N_{I} &= 
N_\text{DM} \cdot N_\ell N_j N_n^{(q)} N_n^{(v)},
\end{align}
which can be also be substantial. So,
\eqref{time:tot} is typically dominated by the $T_\text{proj.}^{(q,v)}$ and $T_{I}$ terms. 

Note that the limit of large $N_{g_\chi} N_{f_S}$, once $T_{K}$ and $T_\text{Tr}$ become the slowest part of an analysis, it may be worthwhile to evaluate the  vector products using GPU rather than CPU processors. It is likely that GPUs could improve substantially upon \eqref{eq:forGPU}.

Compare this $T_\text{tot}$ to the direct, repeated integration of \eqref{eq:rateE}:
\begin{align}
T[\text{\eqref{eq:rateE}}] &= N_{g_\chi} N_{f_S} N_\text{DM} N_\mathcal{R} \cdot T_\text{num.int.}, 
\end{align}
where $T_\text{num.int.}$ is the evaluation time for a single numerical integral. For a relatively easy case with asymmetric but analytic models for $g_\chi$ and $f_s^2$,  Ref.~\cite{Lillard:2023cyy} finds $T_\text{num.int.} \sim 600\, \text{s}$~\footnote{Both versions of the analysis used the same programming language (Python) for the numeric integration and the vector algebra, with similar precision goals of $0.1\%$ in the scattering rate. Both versions of the calculation can be made faster with some optimizations or in other programming languages.}.

\section{Generalization} \label{sec:general}

The vector space integration method works because the scattering rate integrand depends linearly on the functions $g_\chi$ and $f_S^2$. It works especially well because the scattering operator $F_\text{DM}^2 \,\delta(\ldots)$ is block-diagonal in a spherical harmonic basis. 
Similar situations occur in fields well outside of particle physics, so this method is broadly applicable, e.g.~to quantum-mechanical problems~\cite{Kievsky:2008es} or multivariate statistical analyses. 
Any normalizable integral of the form
\begin{align}
S = \int\! d\Pi \, f_1(\vec x_1) f_2(\vec x_2) \ldots f_k(\vec x_k) \, \hat{\mathcal O}(\vec x_1, \ldots, \vec x_k; \vartheta)
\label{genericI}
\end{align}
for operator $\hat{\mathcal O}$ and functions $f_i$ can be expanded using sets of basis functions $\phi^{(i)}$,
\begin{align}
S =\sum_{\phi^{(i)}} \langle \phi^{(1)} | f_1 \rangle \ldots \langle \phi^{(k)} | f_k \rangle \int\! d\Pi \, \phi^{(1)}  \ldots \phi^{(k)} \cdot \hat{\mathcal O}(\vartheta) .
\label{genericR}
\end{align}
Here $\vartheta$ represents any external parameters, e.g.~the DM mass $m_\chi$. 
If a choice of basis functions $\phi^{(i)}$ permits the analytic evaluation of the integral in \eqref{genericR}, then the factorized calculations of $\ket{f_i}$ and the rank~$k$ tensor operator $\mathcal O(\vartheta)$ can be (much) faster than the original \eqref{genericI}, when many versions of each $f_i$ and $\vartheta$ must be tested simultaneously. 
An analogue of the partial rate matrix can be constructed for $S$ if any of the input functions $f_i$ are to be rotated. 
This partial rate tensor, a function of the $d$-dimensional rotation operators $\mathcal R_i \in SO(d)$ acting on the $i$th input function, 
would enable rapid evaluations of $S(\{ \mathcal R_i \})$ under different configurations.

Other applications require not just $S$, but also its derivatives: either with respect to the parameters $\vartheta$, or as functional derivatives of the form $\delta S/\delta f_i$. 
The derivatives $\partial/ \partial\vartheta_i$ can be applied directly to the tensor operator $\hat{\mathcal O}$, sometimes producing closed form analytic expressions. 
In the latter case, the finite-dimensional approximation of the functional derivative (i.e.~replacing $\delta f_i$ with $\ket{ \delta f_i}$) is given by the derivative of \eqref{genericR} with respect to the coefficients $\langle \phi^{(i)} | f_i \rangle$, i.e.:
\begin{align}
\frac{\delta S}{\delta f_i} &\simeq \langle f_1 f_2 \ldots f_{i-1} | \hat{\mathcal O}(\vartheta)  | f_{i+1} \ldots f_{k} \rangle.
\label{newfunctional}
\end{align}

\section{Conclusion}

Using the partial rate matrix $K^{(\ell)}_{mm'}$, the direct detection calculation becomes several orders of magnitude faster. 
In the limit of large $N_\mathcal{R}$, the reduction in computation time is so extreme that an analysis that would otherwise have taken \emph{thousands of cpu-years} can instead be completed on a 2014 MacBook Air in a few hours~\footnote{Taking $T_\text{num.int.} / T_\text{Tr} \sim 10^7$, and the limit where \eqref{time:tot} is dominated by the $N^4 T_\text{Tr}$ term, assuming that tabulated versions of $f_S^2$ and $g_\chi$ or the coefficients in $\ket{f_S^2}$, $\ket{g_\chi}$ are already known.}.
A Python implementation of this method, Vector Spaces for Dark Matter (\texttt{vsdm}), is publicly available~\cite{Lillard_VSDM_2023} and can be installed via:
\begin{align}
\texttt{pip install vsdm} .
\end{align}

There are several immediate applications. The partial rate matrix \eqref{eq:rateTrace} makes it easy to determine the optimal orientations of an anisotropic DM detector, or to determine how variations in $g_\chi$ affect the DM scattering rates. 
The form of \eqref{eq:rateTrace} also provides new information for selecting target materials: for example, to be sensitive to the leading $\ell = 1$ anisotropy in $g_\chi$, the detector material needs to have a large $\ell = 1$ response. Center-symmetric crystals such as trans-stilbene~\cite{Blanco:2021hlm} have support only at even $\ell$, suggesting that even larger modulation amplitudes could be discovered in materials that lack this discrete symmetry.

In the optimistic scenario where direct detection experiments discover evidence for dark matter, the accelerated rate calculation makes it much easier to conduct the statistical analyses needed to extract information about the particle model $(m_\chi, F_\text{DM})$ from the data. To account for the uncertainty in $g_\chi$, such an analysis could be performed on an ensemble of realistic $g_\chi$ models, or in a model-independent way by varying the coefficients of $\ket{g_\chi}$ (in the spirit of e.g.~\cite{Fox:2010bu, Fox:2010bz,Gondolo:2012rs,Kavanagh:2013eya,Feldstein:2014gza,Chen:2021qao,Chen:2022xzi}). 
In the farther future, with large anisotropic direct detection experiments measuring $R(\mathcal R)$ with precision, the experimentally determined values of $K^{(\ell)}_{m m'}$ can be translated into measurements of the DM velocity distribution, in what is sometimes referred to as DM direct detection astronomy~\cite{Peter:2011eu,Peter:2013aha,Lee:2014cpa}.

A relatively small number of coefficients can accurately represent the full 3d function much more concisely than a similarly precise grid of $g_\chi(\vec v)$ or $f_S^2(\vec q, E)$ values. 
In contemporary work this is especially useful for the momentum form factor $f_S^2$, which is often derived numerically. A precise representation of $f_S(\vec q, E)$ may require upwards of $10^9$ grid points, so this input to the rate integrand alone requires many gigabytes of memory. 
Numerically derived $g_\chi$ and $f_S^2$ models can be precisely reconstructed from mere kilobytes of data (as in Fig.~\ref{fig:accuracy}),
making it easy to save, share and apply new results from astrophysics or materials science.
With buy-in from the community, or with moderate individual efforts, a library of $\ket{f_S^2}$ and $\ket{g_\chi}$ coefficients could be assembled for all the detector materials and astrophysical models currently under consideration. Among other benefits, this shared resource would make it much easier to reproduce results from other researchers.

\begin{acknowledgments}

I thank Patrick Draper for collaboration in an early phase of this project.
I thank
Pouya Asadi,
Carlos Blanco,
Matthew Buckley,
David Curtan, 
Valerie Domcke, 
Peizhi Du,
Paddy Fox,
Yonatan Kahn,
David~E. Kaplan,
Graham Kribs,
Mariangela Lisante,
Samuel McDermott,
Robert McGehee,
Surjeet Rajendran,
David Shih,
Dave Soper,
Scott Thomas,
Ken Van Tilberg,
Tien-Tien Yu, 
and Zhengkang Kevin Zhang
for helpful and thought-provoking conversations.
Thanks to Yonatan Kahn for helpful comments on an early draft of this paper.
I thank Ken Van Tilberg for hospitality at the Center for Computational Astrophysics, where some of this work was performed.
This work was supported in part by the U.S. Department of Energy under Grant Number DE-SC0011640.
This work was performed in part at the Aspen Center for Physics, which is supported by National Science Foundation grant PHY-2210452.
This work did not benefit from any access to high-performance computing facilities.

\end{acknowledgments}

\appendix

%
%This supplemental material introduces the spherical wavelet basis functions defined in Ref.~\cite{Lillard:2023cyy}, as well as the technical details that went into the production of Fig.~1. 
%Section~\ref{sec:RI} includes a discussion of the resolution of the identity method. 

\appendix 

\section{Spherical Wavelet Basis Functions} \label{sec:wavelets}

Ref.~\cite{Lillard:2023cyy} introduces orthogonal spherical wavelets $h_{\lambda \mu}(x)$ defined on $x \in [0,1]$, 
\begin{align}
\int_0^1 \! x^2 dx\, h_{\lambda\mu}(x) \, h_{\lambda'\mu'}(x) = \delta_\lambda^{\lambda'} \delta_\mu^{\mu'},
\label{eq:hlm}
\end{align}
for $\lambda = 0, 1, 2, \ldots$, and $ 0 \leq \mu \leq 2^{\lambda} - 1$. Each wavelet is piecewise-constant, 
\begin{align}
{h}_{\lambda \mu} (x) = 
\left\{
\begin{array}{r c l}
+A_{\lambda \mu} && x_1 \leq x <  x_2,
\\[4pt]
-B_{\lambda \mu} &&  x_2 < x \leq x_3 ,
\\[4pt]
0 && \text{otherwise} ,
\end{array}
\right.
\end{align}
where
\begin{align}
\{ x_1, x_2, x_3\} &=  \{ 2^{-\lambda} \mu,\ 2^{-\lambda}( \mu + \tfrac{1}{2}) ,\ 2^{-\lambda}(\mu+1) \} ,
\label{def:x123}
\end{align}
and 
\begin{align}
A_{\lambda \mu} &= \sqrt{ \frac{3}{x_3^3 - x_1^3} \frac{x_3^3 - x_2^3 }{x_2^3 - x_1^3} } ,
\\
B_{\lambda \mu} &= \sqrt{ \frac{3}{x_3^3 - x_1^3} \frac{x_2^3 - x_1^3}{x_3^3 - x_2^3 } }  .
\end{align}
To match the $\ket{n \ell m }$ notation, $(\lambda, \mu)$ is mapped onto a single integer-valued index $n$,
\begin{align}
n(\lambda \geq 0) &\equiv 2^\lambda + \mu.
\end{align}
A constant wavelet (``$\lambda=-1$'')  completes the basis:
\begin{align}
h_{n=0}(0 \leq x \leq 1) &\equiv \sqrt{3} .
\end{align}
Three dimensional functions $f(\vec u)$ of finite range $|\vec u| \leq u_\text{max}$ can be expanded in a spherical wavelet-harmonic basis,
\begin{align}
\phi_{n \ell m}(\vec u) &\equiv r_n(u) \, Y_{\ell m}(\hat u), 
\\
r_n(u) &= h_{n}(u/u_\text{max}) .
\end{align} 
These $\phi_{n \ell m}$ are mutually orthogonal and complete.

For 4d functions of $(\vec q, E)$, the wavelet-harmonic basis can be expanded to include energy basis functions $R_j(E)$, 
\begin{align}
\varphi_{j n \ell m}(\vec q, E) &= R_j(E) \, \phi_{n \ell m}(\vec q) , 
\end{align}
for example using Haar wavelets: 
\begin{align}
H_{\lambda\mu}(x) &\equiv \left\{ \begin{array}{c c l}
	+2^{\lambda/2},  && 2^\lambda x \in [ \mu ,\,  \mu + \frac{1}{2}  ) , \\
	-2^{\lambda/2},  && 2^\lambda x \in (  \mu + \frac{1}{2}  ,\,  \mu+ 1] , \end{array} \right.
\end{align}
with $H_{-1}(x) \equiv 1$ completing the basis. In the case of discrete final state energies $\Delta E_j$, as in Fig.~1, $R_j(E) \rightarrow \delta(E - \Delta E_j)$. 

Spherical wavelet basis functions are easily derived in higher dimensions, e.g.~for generalizations of Sec.~IV
in cases where the operator $\mathcal O$ in Eq.~(25) is a function of $p$-dimensional dot products $\vec x_i \cdot \vec x_j$. 
The spherical harmonics $Y_{\ell m}$ are replaced by hyperspherical harmonics (i.e.~Gegenbauer functions), while the values of $A_{\lambda \mu}$ and $B_{\lambda \mu}$ are trivially modified to ensure that the hyperspherical wavelets $h_{\lambda \mu}$ are mutually orthogonal under the weighting function $x^2 \rightarrow x^{p-1}$ in \eqref{eq:hlm}. This procedure also works for 2d polar wavelets.

\section{Particle In A Box} \label{appx:B}

To quantify the rates of convergence of the basis function expansion, Fig.~1 uses a simple particle-in-a-box model to generate a set of analytic detector response functions, borrowed from the companion paper~\cite{Lillard:2023cyy}. 
Here $f_s^2(\vec q)$ is derived from initial and final state wavefunctions satisfying
\begin{align}
\Psi_{\vec n}(\vec r) = \frac{2^{3/2} }{\sqrt{V_\text{cell}}} \sin\frac{\pi n_x x}{L_x}  \sin\frac{\pi n_y y}{L_y}  \sin\frac{\pi n_z z}{L_z} ,
\label{eq:wavebox}
\end{align}
where $\vec n = (n_x, n_y, n_z)$ labels the energy eigenstates of the system, with energies 
\begin{align}
\Delta E_{\vec n} &= \frac{\pi^2 }{2m} \left( \frac{n_x^2 - 1}{L_x^2} + \frac{n_y^2 - 1}{L_y^2 } + \frac{n_z^2 -1}{L_z^2} \right) .
\end{align}
Here $m$ is the mass of the particle in the box, a free parameter in this model. 
The ground state is $\vec n = (1, 1, 1)$, and the first excited state sets $n_i = 2$ for the $i = x, y, z$ corresponding to the largest value of $L_i$. 

From 
\begin{align}
f_s(\vec q) &= \left\langle \Psi_{\vec n}(\vec r) \Big| e^{i \vec q \cdot \vec r } \Big| \Psi_0(\vec r) \right\rangle ,
\end{align}
it can be shown that the momentum form factor $f_s^2(\vec q) \equiv |f_s|^2$ for a transition from the ground state to the excited state
$\vec n$ is given by:
\begin{widetext}
\begin{align}
f_s^2(\vec q) &= \!\!\prod_{j = x, y, z} \left[ \frac{\sinc\left( \frac{|q_j L_j| - \pi (n_j - 1) }{2} \right) }{1 + \pi(n_j - 1)/|q_j L_j|  }  +  \frac{\sinc\left( \frac{|q_j L_j| - \pi (n_j + 1) }{2} \right) }{1 + \pi(n_j + 1)/|q_j L_j|  }  \right]^2. 
\label{fS2:box}
\end{align} 
\end{widetext}
The box is taken to have dimensions
\begin{align}
\vec L = (4, 7, 10) \times (\alpha m_e)^{-1}, 
\end{align}
so that $\vec n = (1, 1, 2)$ is the first excited state. 

In the Fig.~1 demonstration, it is the $\vec n = (3, 2, 1)$ excited state that is shown.
This $f_s^2(\vec q)$ has maxima at $\vec q = (\pm 8.73 \, \text{keV},  \pm 3.25\, \text{keV}, 0)$, where $f_s^2(\vec q) = 0.198$. 
The nodes, where $f_s^2(\vec q)$ vanishes, are located on the $q_x = 0$ and $q_y = 0$ planes.
For further details, see Ref.~\cite{Lillard:2023cyy}.

\bibliography{dmvec_refs}

\begin{thebibliography}{62}
\expandafter\ifx\csname natexlab\endcsname\relax\def\natexlab#1{#1}\fi
\expandafter\ifx\csname bibnamefont\endcsname\relax
  \def\bibnamefont#1{#1}\fi
\expandafter\ifx\csname bibfnamefont\endcsname\relax
  \def\bibfnamefont#1{#1}\fi
\expandafter\ifx\csname citenamefont\endcsname\relax
  \def\citenamefont#1{#1}\fi
\expandafter\ifx\csname url\endcsname\relax
  \def\url#1{\texttt{#1}}\fi
\expandafter\ifx\csname urlprefix\endcsname\relax\def\urlprefix{URL }\fi
\providecommand{\bibinfo}[2]{#2}
\providecommand{\eprint}[2][]{\url{#2}}

\bibitem[{\citenamefont{Agnese et~al.}(2018)}]{SuperCDMS:2018mne}
\bibinfo{author}{\bibfnamefont{R.}~\bibnamefont{Agnese}} \bibnamefont{et~al.}
  (\bibinfo{collaboration}{SuperCDMS}), \bibinfo{journal}{Phys. Rev. Lett.}
  \textbf{\bibinfo{volume}{121}}, \bibinfo{pages}{051301}
  (\bibinfo{year}{2018}), \bibinfo{note}{[Erratum: Phys.Rev.Lett. 122, 069901
  (2019)]}, \eprint{1804.10697}.

\bibitem[{\citenamefont{Aguilar-Arevalo et~al.}(2019)}]{DAMIC:2019dcn}
\bibinfo{author}{\bibfnamefont{A.}~\bibnamefont{Aguilar-Arevalo}}
  \bibnamefont{et~al.} (\bibinfo{collaboration}{DAMIC}),
  \bibinfo{journal}{Phys. Rev. Lett.} \textbf{\bibinfo{volume}{123}},
  \bibinfo{pages}{181802} (\bibinfo{year}{2019}), \eprint{1907.12628}.

\bibitem[{\citenamefont{Arnaud et~al.}(2020)}]{EDELWEISS:2020fxc}
\bibinfo{author}{\bibfnamefont{Q.}~\bibnamefont{Arnaud}} \bibnamefont{et~al.}
  (\bibinfo{collaboration}{EDELWEISS}), \bibinfo{journal}{Phys. Rev. Lett.}
  \textbf{\bibinfo{volume}{125}}, \bibinfo{pages}{141301}
  (\bibinfo{year}{2020}), \eprint{2003.01046}.

\bibitem[{\citenamefont{Barak et~al.}(2020)}]{SENSEI:2020dpa}
\bibinfo{author}{\bibfnamefont{L.}~\bibnamefont{Barak}} \bibnamefont{et~al.}
  (\bibinfo{collaboration}{SENSEI}), \bibinfo{journal}{Phys. Rev. Lett.}
  \textbf{\bibinfo{volume}{125}}, \bibinfo{pages}{171802}
  (\bibinfo{year}{2020}), \eprint{2004.11378}.

\bibitem[{\citenamefont{Arnquist et~al.}(2023)}]{DAMIC-M:2023gxo}
\bibinfo{author}{\bibfnamefont{I.}~\bibnamefont{Arnquist}} \bibnamefont{et~al.}
  (\bibinfo{collaboration}{DAMIC-M}), \bibinfo{journal}{Phys. Rev. Lett.}
  \textbf{\bibinfo{volume}{130}}, \bibinfo{pages}{171003}
  (\bibinfo{year}{2023}), \eprint{2302.02372}.

\bibitem[{\citenamefont{Blanco et~al.}(2020)\citenamefont{Blanco, Collar, Kahn,
  and Lillard}}]{Blanco:2019lrf}
\bibinfo{author}{\bibfnamefont{C.}~\bibnamefont{Blanco}},
  \bibinfo{author}{\bibfnamefont{J.~I.} \bibnamefont{Collar}},
  \bibinfo{author}{\bibfnamefont{Y.}~\bibnamefont{Kahn}}, \bibnamefont{and}
  \bibinfo{author}{\bibfnamefont{B.}~\bibnamefont{Lillard}},
  \bibinfo{journal}{Phys. Rev. D} \textbf{\bibinfo{volume}{101}},
  \bibinfo{pages}{056001} (\bibinfo{year}{2020}), \eprint{1912.02822}.

\bibitem[{\citenamefont{Blanco et~al.}(2021)\citenamefont{Blanco, Kahn,
  Lillard, and McDermott}}]{Blanco:2021hlm}
\bibinfo{author}{\bibfnamefont{C.}~\bibnamefont{Blanco}},
  \bibinfo{author}{\bibfnamefont{Y.}~\bibnamefont{Kahn}},
  \bibinfo{author}{\bibfnamefont{B.}~\bibnamefont{Lillard}}, \bibnamefont{and}
  \bibinfo{author}{\bibfnamefont{S.~D.} \bibnamefont{McDermott}},
  \bibinfo{journal}{Phys. Rev. D} \textbf{\bibinfo{volume}{104}},
  \bibinfo{pages}{036011} (\bibinfo{year}{2021}), \eprint{2103.08601}.

\bibitem[{\citenamefont{Blanco et~al.}(2022)\citenamefont{Blanco, Harris, Kahn,
  Lillard, and P\'erez-R\'\i{}os}}]{Blanco:2022pkt}
\bibinfo{author}{\bibfnamefont{C.}~\bibnamefont{Blanco}},
  \bibinfo{author}{\bibfnamefont{I.}~\bibnamefont{Harris}},
  \bibinfo{author}{\bibfnamefont{Y.}~\bibnamefont{Kahn}},
  \bibinfo{author}{\bibfnamefont{B.}~\bibnamefont{Lillard}}, \bibnamefont{and}
  \bibinfo{author}{\bibfnamefont{J.}~\bibnamefont{P\'erez-R\'\i{}os}},
  \bibinfo{journal}{Phys. Rev. D} \textbf{\bibinfo{volume}{106}},
  \bibinfo{pages}{115015} (\bibinfo{year}{2022}), \eprint{2208.09002}.

\bibitem[{\citenamefont{Hochberg et~al.}(2017)\citenamefont{Hochberg, Kahn,
  Lisanti, Tully, and Zurek}}]{Hochberg:2016ntt}
\bibinfo{author}{\bibfnamefont{Y.}~\bibnamefont{Hochberg}},
  \bibinfo{author}{\bibfnamefont{Y.}~\bibnamefont{Kahn}},
  \bibinfo{author}{\bibfnamefont{M.}~\bibnamefont{Lisanti}},
  \bibinfo{author}{\bibfnamefont{C.~G.} \bibnamefont{Tully}}, \bibnamefont{and}
  \bibinfo{author}{\bibfnamefont{K.~M.} \bibnamefont{Zurek}},
  \bibinfo{journal}{Phys. Lett. B} \textbf{\bibinfo{volume}{772}},
  \bibinfo{pages}{239} (\bibinfo{year}{2017}), \eprint{1606.08849}.

\bibitem[{\citenamefont{Budnik et~al.}(2018)\citenamefont{Budnik, Chesnovsky,
  Slone, and Volansky}}]{Budnik:2017sbu}
\bibinfo{author}{\bibfnamefont{R.}~\bibnamefont{Budnik}},
  \bibinfo{author}{\bibfnamefont{O.}~\bibnamefont{Chesnovsky}},
  \bibinfo{author}{\bibfnamefont{O.}~\bibnamefont{Slone}}, \bibnamefont{and}
  \bibinfo{author}{\bibfnamefont{T.}~\bibnamefont{Volansky}},
  \bibinfo{journal}{Phys. Lett. B} \textbf{\bibinfo{volume}{782}},
  \bibinfo{pages}{242} (\bibinfo{year}{2018}), \eprint{1705.03016}.

\bibitem[{\citenamefont{Hochberg et~al.}(2018)\citenamefont{Hochberg, Kahn,
  Lisanti, Zurek, Grushin, Ilan, Griffin, Liu, Weber, and
  Neaton}}]{Hochberg:2017wce}
\bibinfo{author}{\bibfnamefont{Y.}~\bibnamefont{Hochberg}},
  \bibinfo{author}{\bibfnamefont{Y.}~\bibnamefont{Kahn}},
  \bibinfo{author}{\bibfnamefont{M.}~\bibnamefont{Lisanti}},
  \bibinfo{author}{\bibfnamefont{K.~M.} \bibnamefont{Zurek}},
  \bibinfo{author}{\bibfnamefont{A.~G.} \bibnamefont{Grushin}},
  \bibinfo{author}{\bibfnamefont{R.}~\bibnamefont{Ilan}},
  \bibinfo{author}{\bibfnamefont{S.~M.} \bibnamefont{Griffin}},
  \bibinfo{author}{\bibfnamefont{Z.-F.} \bibnamefont{Liu}},
  \bibinfo{author}{\bibfnamefont{S.~F.} \bibnamefont{Weber}}, \bibnamefont{and}
  \bibinfo{author}{\bibfnamefont{J.~B.} \bibnamefont{Neaton}},
  \bibinfo{journal}{Phys. Rev. D} \textbf{\bibinfo{volume}{97}},
  \bibinfo{pages}{015004} (\bibinfo{year}{2018}), \eprint{1708.08929}.

\bibitem[{\citenamefont{Coskuner et~al.}(2021)\citenamefont{Coskuner,
  Mitridate, Olivares, and Zurek}}]{Coskuner:2019odd}
\bibinfo{author}{\bibfnamefont{A.}~\bibnamefont{Coskuner}},
  \bibinfo{author}{\bibfnamefont{A.}~\bibnamefont{Mitridate}},
  \bibinfo{author}{\bibfnamefont{A.}~\bibnamefont{Olivares}}, \bibnamefont{and}
  \bibinfo{author}{\bibfnamefont{K.~M.} \bibnamefont{Zurek}},
  \bibinfo{journal}{Phys. Rev. D} \textbf{\bibinfo{volume}{103}},
  \bibinfo{pages}{016006} (\bibinfo{year}{2021}), \eprint{1909.09170}.

\bibitem[{\citenamefont{Geilhufe et~al.}(2020)\citenamefont{Geilhufe,
  Kahlhoefer, and Winkler}}]{Geilhufe:2019ndy}
\bibinfo{author}{\bibfnamefont{R.~M.} \bibnamefont{Geilhufe}},
  \bibinfo{author}{\bibfnamefont{F.}~\bibnamefont{Kahlhoefer}},
  \bibnamefont{and} \bibinfo{author}{\bibfnamefont{M.~W.}
  \bibnamefont{Winkler}}, \bibinfo{journal}{Phys. Rev. D}
  \textbf{\bibinfo{volume}{101}}, \bibinfo{pages}{055005}
  (\bibinfo{year}{2020}), \eprint{1910.02091}.

\bibitem[{\citenamefont{Griffin
  et~al.}(2021{\natexlab{a}})\citenamefont{Griffin, Hochberg, Inzani, Kurinsky,
  Lin, and Chin}}]{Griffin:2020lgd}
\bibinfo{author}{\bibfnamefont{S.~M.} \bibnamefont{Griffin}},
  \bibinfo{author}{\bibfnamefont{Y.}~\bibnamefont{Hochberg}},
  \bibinfo{author}{\bibfnamefont{K.}~\bibnamefont{Inzani}},
  \bibinfo{author}{\bibfnamefont{N.}~\bibnamefont{Kurinsky}},
  \bibinfo{author}{\bibfnamefont{T.}~\bibnamefont{Lin}}, \bibnamefont{and}
  \bibinfo{author}{\bibfnamefont{T.}~\bibnamefont{Chin}},
  \bibinfo{journal}{Phys. Rev. D} \textbf{\bibinfo{volume}{103}},
  \bibinfo{pages}{075002} (\bibinfo{year}{2021}{\natexlab{a}}),
  \eprint{2008.08560}.

\bibitem[{\citenamefont{Coskuner et~al.}(2022)\citenamefont{Coskuner, Trickle,
  Zhang, and Zurek}}]{Coskuner:2021qxo}
\bibinfo{author}{\bibfnamefont{A.}~\bibnamefont{Coskuner}},
  \bibinfo{author}{\bibfnamefont{T.}~\bibnamefont{Trickle}},
  \bibinfo{author}{\bibfnamefont{Z.}~\bibnamefont{Zhang}}, \bibnamefont{and}
  \bibinfo{author}{\bibfnamefont{K.~M.} \bibnamefont{Zurek}},
  \bibinfo{journal}{Phys. Rev. D} \textbf{\bibinfo{volume}{105}},
  \bibinfo{pages}{015010} (\bibinfo{year}{2022}), \eprint{2102.09567}.

\bibitem[{\citenamefont{Boyd et~al.}(2022)\citenamefont{Boyd, Hochberg, Kahn,
  Kramer, Kurinsky, Lehmann, and Yu}}]{Boyd:2022tcn}
\bibinfo{author}{\bibfnamefont{C.}~\bibnamefont{Boyd}},
  \bibinfo{author}{\bibfnamefont{Y.}~\bibnamefont{Hochberg}},
  \bibinfo{author}{\bibfnamefont{Y.}~\bibnamefont{Kahn}},
  \bibinfo{author}{\bibfnamefont{E.~D.} \bibnamefont{Kramer}},
  \bibinfo{author}{\bibfnamefont{N.}~\bibnamefont{Kurinsky}},
  \bibinfo{author}{\bibfnamefont{B.~V.} \bibnamefont{Lehmann}},
  \bibnamefont{and} \bibinfo{author}{\bibfnamefont{T.~C.} \bibnamefont{Yu}}
  (\bibinfo{year}{2022}), \eprint{2212.04505}.

\bibitem[{\citenamefont{Catena et~al.}(2023{\natexlab{a}})\citenamefont{Catena,
  Emken, Matas, Spaldin, and Urdshals}}]{Catena:2023qkj}
\bibinfo{author}{\bibfnamefont{R.}~\bibnamefont{Catena}},
  \bibinfo{author}{\bibfnamefont{T.}~\bibnamefont{Emken}},
  \bibinfo{author}{\bibfnamefont{M.}~\bibnamefont{Matas}},
  \bibinfo{author}{\bibfnamefont{N.~A.} \bibnamefont{Spaldin}},
  \bibnamefont{and} \bibinfo{author}{\bibfnamefont{E.}~\bibnamefont{Urdshals}}
  (\bibinfo{year}{2023}{\natexlab{a}}), \eprint{2303.15497}.

\bibitem[{\citenamefont{Catena et~al.}(2023{\natexlab{b}})\citenamefont{Catena,
  Emken, Matas, Spaldin, and Urdshals}}]{Catena:2023awl}
\bibinfo{author}{\bibfnamefont{R.}~\bibnamefont{Catena}},
  \bibinfo{author}{\bibfnamefont{T.}~\bibnamefont{Emken}},
  \bibinfo{author}{\bibfnamefont{M.}~\bibnamefont{Matas}},
  \bibinfo{author}{\bibfnamefont{N.~A.} \bibnamefont{Spaldin}},
  \bibnamefont{and} \bibinfo{author}{\bibfnamefont{E.}~\bibnamefont{Urdshals}}
  (\bibinfo{year}{2023}{\natexlab{b}}), \eprint{2303.15509}.

\bibitem[{\citenamefont{Aprile et~al.}(2018)}]{XENON:2018voc}
\bibinfo{author}{\bibfnamefont{E.}~\bibnamefont{Aprile}} \bibnamefont{et~al.}
  (\bibinfo{collaboration}{XENON}), \bibinfo{journal}{Phys. Rev. Lett.}
  \textbf{\bibinfo{volume}{121}}, \bibinfo{pages}{111302}
  (\bibinfo{year}{2018}), \eprint{1805.12562}.

\bibitem[{\citenamefont{Cheng et~al.}(2021)}]{PandaX-II:2021nsg}
\bibinfo{author}{\bibfnamefont{C.}~\bibnamefont{Cheng}} \bibnamefont{et~al.}
  (\bibinfo{collaboration}{PandaX-II}), \bibinfo{journal}{Phys. Rev. Lett.}
  \textbf{\bibinfo{volume}{126}}, \bibinfo{pages}{211803}
  (\bibinfo{year}{2021}), \eprint{2101.07479}.

\bibitem[{\citenamefont{Aalbers et~al.}(2022)}]{LZ:2022ufs}
\bibinfo{author}{\bibfnamefont{J.}~\bibnamefont{Aalbers}} \bibnamefont{et~al.}
  (\bibinfo{collaboration}{LZ}) (\bibinfo{year}{2022}), \eprint{2207.03764}.

\bibitem[{\citenamefont{Aprile et~al.}(2023)}]{XENON:2023sxq}
\bibinfo{author}{\bibfnamefont{E.}~\bibnamefont{Aprile}} \bibnamefont{et~al.}
  (\bibinfo{collaboration}{XENON}) (\bibinfo{year}{2023}), \eprint{2303.14729}.

\bibitem[{\citenamefont{Agnes et~al.}(2018)}]{DarkSide:2018ppu}
\bibinfo{author}{\bibfnamefont{P.}~\bibnamefont{Agnes}} \bibnamefont{et~al.}
  (\bibinfo{collaboration}{DarkSide}), \bibinfo{journal}{Phys. Rev. Lett.}
  \textbf{\bibinfo{volume}{121}}, \bibinfo{pages}{111303}
  (\bibinfo{year}{2018}), \eprint{1802.06998}.

\bibitem[{\citenamefont{Aprile et~al.}(2022)}]{XENON:2022ltv}
\bibinfo{author}{\bibfnamefont{E.}~\bibnamefont{Aprile}} \bibnamefont{et~al.}
  (\bibinfo{collaboration}{XENON}), \bibinfo{journal}{Phys. Rev. Lett.}
  \textbf{\bibinfo{volume}{129}}, \bibinfo{pages}{161805}
  (\bibinfo{year}{2022}), \eprint{2207.11330}.

\bibitem[{\citenamefont{Lewin and Smith}(1996)}]{Lewin:1995rx}
\bibinfo{author}{\bibfnamefont{J.~D.} \bibnamefont{Lewin}} \bibnamefont{and}
  \bibinfo{author}{\bibfnamefont{P.~F.} \bibnamefont{Smith}},
  \bibinfo{journal}{Astropart. Phys.} \textbf{\bibinfo{volume}{6}},
  \bibinfo{pages}{87} (\bibinfo{year}{1996}).

\bibitem[{\citenamefont{Drukier et~al.}(1986)\citenamefont{Drukier, Freese, and
  Spergel}}]{Drukier:1986tm}
\bibinfo{author}{\bibfnamefont{A.~K.} \bibnamefont{Drukier}},
  \bibinfo{author}{\bibfnamefont{K.}~\bibnamefont{Freese}}, \bibnamefont{and}
  \bibinfo{author}{\bibfnamefont{D.~N.} \bibnamefont{Spergel}},
  \bibinfo{journal}{Phys. Rev. D} \textbf{\bibinfo{volume}{33}},
  \bibinfo{pages}{3495} (\bibinfo{year}{1986}).

\bibitem[{\citenamefont{Freese et~al.}(1988)\citenamefont{Freese, Frieman, and
  Gould}}]{Freese:1987wu}
\bibinfo{author}{\bibfnamefont{K.}~\bibnamefont{Freese}},
  \bibinfo{author}{\bibfnamefont{J.~A.} \bibnamefont{Frieman}},
  \bibnamefont{and} \bibinfo{author}{\bibfnamefont{A.}~\bibnamefont{Gould}},
  \bibinfo{journal}{Phys. Rev. D} \textbf{\bibinfo{volume}{37}},
  \bibinfo{pages}{3388} (\bibinfo{year}{1988}).

\bibitem[{\citenamefont{Lee et~al.}(2015)\citenamefont{Lee, Lisanti,
  Mishra-Sharma, and Safdi}}]{Lee:2015qva}
\bibinfo{author}{\bibfnamefont{S.~K.} \bibnamefont{Lee}},
  \bibinfo{author}{\bibfnamefont{M.}~\bibnamefont{Lisanti}},
  \bibinfo{author}{\bibfnamefont{S.}~\bibnamefont{Mishra-Sharma}},
  \bibnamefont{and} \bibinfo{author}{\bibfnamefont{B.~R.} \bibnamefont{Safdi}},
  \bibinfo{journal}{Phys. Rev. D} \textbf{\bibinfo{volume}{92}},
  \bibinfo{pages}{083517} (\bibinfo{year}{2015}), \eprint{1508.07361}.

\bibitem[{\citenamefont{Henderson}(1839)}]{10.1093/mnras/4.19.168}
\bibinfo{author}{\bibnamefont{Henderson}}, \bibinfo{journal}{Monthly Notices of
  the Royal Astronomical Society} \textbf{\bibinfo{volume}{4}},
  \bibinfo{pages}{168} (\bibinfo{year}{1839}), ISSN \bibinfo{issn}{0035-8711}.

\bibitem[{\citenamefont{Diemand et~al.}(2008)\citenamefont{Diemand, Kuhlen,
  Madau, Zemp, Moore, Potter, and Stadel}}]{Diemand:2008in}
\bibinfo{author}{\bibfnamefont{J.}~\bibnamefont{Diemand}},
  \bibinfo{author}{\bibfnamefont{M.}~\bibnamefont{Kuhlen}},
  \bibinfo{author}{\bibfnamefont{P.}~\bibnamefont{Madau}},
  \bibinfo{author}{\bibfnamefont{M.}~\bibnamefont{Zemp}},
  \bibinfo{author}{\bibfnamefont{B.}~\bibnamefont{Moore}},
  \bibinfo{author}{\bibfnamefont{D.}~\bibnamefont{Potter}}, \bibnamefont{and}
  \bibinfo{author}{\bibfnamefont{J.}~\bibnamefont{Stadel}},
  \bibinfo{journal}{Nature} \textbf{\bibinfo{volume}{454}},
  \bibinfo{pages}{735} (\bibinfo{year}{2008}), \eprint{0805.1244}.

\bibitem[{\citenamefont{Klypin et~al.}(2011)\citenamefont{Klypin,
  Trujillo-Gomez, and Primack}}]{Klypin:2010qw}
\bibinfo{author}{\bibfnamefont{A.}~\bibnamefont{Klypin}},
  \bibinfo{author}{\bibfnamefont{S.}~\bibnamefont{Trujillo-Gomez}},
  \bibnamefont{and} \bibinfo{author}{\bibfnamefont{J.}~\bibnamefont{Primack}},
  \bibinfo{journal}{Astrophys. J.} \textbf{\bibinfo{volume}{740}},
  \bibinfo{pages}{102} (\bibinfo{year}{2011}), \eprint{1002.3660}.

\bibitem[{\citenamefont{Guedes et~al.}(2011)\citenamefont{Guedes, Callegari,
  Madau, and Mayer}}]{Guedes:2011ux}
\bibinfo{author}{\bibfnamefont{J.}~\bibnamefont{Guedes}},
  \bibinfo{author}{\bibfnamefont{S.}~\bibnamefont{Callegari}},
  \bibinfo{author}{\bibfnamefont{P.}~\bibnamefont{Madau}}, \bibnamefont{and}
  \bibinfo{author}{\bibfnamefont{L.}~\bibnamefont{Mayer}},
  \bibinfo{journal}{Astrophys. J.} \textbf{\bibinfo{volume}{742}},
  \bibinfo{pages}{76} (\bibinfo{year}{2011}), \eprint{1103.6030}.

\bibitem[{\citenamefont{Hopkins et~al.}(2018)}]{Hopkins:2017ycn}
\bibinfo{author}{\bibfnamefont{P.~F.} \bibnamefont{Hopkins}}
  \bibnamefont{et~al.}, \bibinfo{journal}{Mon. Not. Roy. Astron. Soc.}
  \textbf{\bibinfo{volume}{480}}, \bibinfo{pages}{800} (\bibinfo{year}{2018}),
  \eprint{1702.06148}.

\bibitem[{\citenamefont{Kuhlen et~al.}(2012)\citenamefont{Kuhlen, Lisanti, and
  Spergel}}]{Kuhlen:2012fz}
\bibinfo{author}{\bibfnamefont{M.}~\bibnamefont{Kuhlen}},
  \bibinfo{author}{\bibfnamefont{M.}~\bibnamefont{Lisanti}}, \bibnamefont{and}
  \bibinfo{author}{\bibfnamefont{D.~N.} \bibnamefont{Spergel}},
  \bibinfo{journal}{Phys. Rev. D} \textbf{\bibinfo{volume}{86}},
  \bibinfo{pages}{063505} (\bibinfo{year}{2012}), \eprint{1202.0007}.

\bibitem[{\citenamefont{Necib et~al.}(2018)\citenamefont{Necib, Lisanti,
  Garrison-Kimmel, Wetzel, Sanderson, Hopkins, Faucher-Gigu\`ere, and
  Kere\v{s}}}]{Necib:2018igl}
\bibinfo{author}{\bibfnamefont{L.}~\bibnamefont{Necib}},
  \bibinfo{author}{\bibfnamefont{M.}~\bibnamefont{Lisanti}},
  \bibinfo{author}{\bibfnamefont{S.}~\bibnamefont{Garrison-Kimmel}},
  \bibinfo{author}{\bibfnamefont{A.}~\bibnamefont{Wetzel}},
  \bibinfo{author}{\bibfnamefont{R.}~\bibnamefont{Sanderson}},
  \bibinfo{author}{\bibfnamefont{P.~F.} \bibnamefont{Hopkins}},
  \bibinfo{author}{\bibfnamefont{C.-A.} \bibnamefont{Faucher-Gigu\`ere}},
  \bibnamefont{and} \bibinfo{author}{\bibfnamefont{D.}~\bibnamefont{Kere\v{s}}}
  (\bibinfo{year}{2018}), \eprint{1810.12301}.

\bibitem[{\citenamefont{Riley et~al.}(2019)}]{Riley:2018lbh}
\bibinfo{author}{\bibfnamefont{A.~H.} \bibnamefont{Riley}}
  \bibnamefont{et~al.}, \bibinfo{journal}{Mon. Not. Roy. Astron. Soc.}
  \textbf{\bibinfo{volume}{486}}, \bibinfo{pages}{2679} (\bibinfo{year}{2019}),
  \eprint{1810.10645}.

\bibitem[{\citenamefont{Wu et~al.}(2019)\citenamefont{Wu, Freese, Kelso,
  Stengel, and Valluri}}]{Wu:2019nhd}
\bibinfo{author}{\bibfnamefont{Y.}~\bibnamefont{Wu}},
  \bibinfo{author}{\bibfnamefont{K.}~\bibnamefont{Freese}},
  \bibinfo{author}{\bibfnamefont{C.}~\bibnamefont{Kelso}},
  \bibinfo{author}{\bibfnamefont{P.}~\bibnamefont{Stengel}}, \bibnamefont{and}
  \bibinfo{author}{\bibfnamefont{M.}~\bibnamefont{Valluri}},
  \bibinfo{journal}{JCAP} \textbf{\bibinfo{volume}{10}}, \bibinfo{pages}{034}
  (\bibinfo{year}{2019}), \eprint{1904.04781}.

\bibitem[{\citenamefont{Radick et~al.}(2021)\citenamefont{Radick, Taki, and
  Yu}}]{Radick:2020qip}
\bibinfo{author}{\bibfnamefont{A.}~\bibnamefont{Radick}},
  \bibinfo{author}{\bibfnamefont{A.-M.} \bibnamefont{Taki}}, \bibnamefont{and}
  \bibinfo{author}{\bibfnamefont{T.-T.} \bibnamefont{Yu}},
  \bibinfo{journal}{JCAP} \textbf{\bibinfo{volume}{02}}, \bibinfo{pages}{004}
  (\bibinfo{year}{2021}), \eprint{2011.02493}.

\bibitem[{\citenamefont{Maity et~al.}(2021)\citenamefont{Maity, Ray, and
  Sarkar}}]{Maity:2020wic}
\bibinfo{author}{\bibfnamefont{T.~N.} \bibnamefont{Maity}},
  \bibinfo{author}{\bibfnamefont{T.~S.} \bibnamefont{Ray}}, \bibnamefont{and}
  \bibinfo{author}{\bibfnamefont{S.}~\bibnamefont{Sarkar}},
  \bibinfo{journal}{Eur. Phys. J. C} \textbf{\bibinfo{volume}{81}},
  \bibinfo{pages}{1005} (\bibinfo{year}{2021}), \eprint{2011.12896}.

\bibitem[{\citenamefont{Buckley et~al.}(2022)\citenamefont{Buckley, Lim,
  Putney, and Shih}}]{Buckley:2022tjy}
\bibinfo{author}{\bibfnamefont{M.~R.} \bibnamefont{Buckley}},
  \bibinfo{author}{\bibfnamefont{S.~H.} \bibnamefont{Lim}},
  \bibinfo{author}{\bibfnamefont{E.}~\bibnamefont{Putney}}, \bibnamefont{and}
  \bibinfo{author}{\bibfnamefont{D.}~\bibnamefont{Shih}}
  (\bibinfo{year}{2022}), \eprint{2205.01129}.

\bibitem[{\citenamefont{Maity and Laha}(2023)}]{Maity:2022enp}
\bibinfo{author}{\bibfnamefont{T.~N.} \bibnamefont{Maity}} \bibnamefont{and}
  \bibinfo{author}{\bibfnamefont{R.}~\bibnamefont{Laha}},
  \bibinfo{journal}{JHEP} \textbf{\bibinfo{volume}{02}}, \bibinfo{pages}{200}
  (\bibinfo{year}{2023}), \eprint{2208.14471}.

\bibitem[{\citenamefont{Knapen et~al.}(2021)\citenamefont{Knapen, Kozaczuk, and
  Lin}}]{Knapen:2021run}
\bibinfo{author}{\bibfnamefont{S.}~\bibnamefont{Knapen}},
  \bibinfo{author}{\bibfnamefont{J.}~\bibnamefont{Kozaczuk}}, \bibnamefont{and}
  \bibinfo{author}{\bibfnamefont{T.}~\bibnamefont{Lin}},
  \bibinfo{journal}{Phys. Rev. D} \textbf{\bibinfo{volume}{104}},
  \bibinfo{pages}{015031} (\bibinfo{year}{2021}), \eprint{2101.08275}.

\bibitem[{\citenamefont{Knapen et~al.}(2022)\citenamefont{Knapen, Kozaczuk, and
  Lin}}]{Knapen:2021bwg}
\bibinfo{author}{\bibfnamefont{S.}~\bibnamefont{Knapen}},
  \bibinfo{author}{\bibfnamefont{J.}~\bibnamefont{Kozaczuk}}, \bibnamefont{and}
  \bibinfo{author}{\bibfnamefont{T.}~\bibnamefont{Lin}},
  \bibinfo{journal}{Phys. Rev. D} \textbf{\bibinfo{volume}{105}},
  \bibinfo{pages}{015014} (\bibinfo{year}{2022}), \eprint{2104.12786}.

\bibitem[{\citenamefont{Griffin
  et~al.}(2021{\natexlab{b}})\citenamefont{Griffin, Inzani, Trickle, Zhang, and
  Zurek}}]{Griffin:2021znd}
\bibinfo{author}{\bibfnamefont{S.~M.} \bibnamefont{Griffin}},
  \bibinfo{author}{\bibfnamefont{K.}~\bibnamefont{Inzani}},
  \bibinfo{author}{\bibfnamefont{T.}~\bibnamefont{Trickle}},
  \bibinfo{author}{\bibfnamefont{Z.}~\bibnamefont{Zhang}}, \bibnamefont{and}
  \bibinfo{author}{\bibfnamefont{K.~M.} \bibnamefont{Zurek}},
  \bibinfo{journal}{Phys. Rev. D} \textbf{\bibinfo{volume}{104}},
  \bibinfo{pages}{095015} (\bibinfo{year}{2021}{\natexlab{b}}),
  \eprint{2105.05253}.

\bibitem[{\citenamefont{Lillard}(2025)}]{Lillard:2023cyy}
\bibinfo{author}{\bibfnamefont{B.}~\bibnamefont{Lillard}},
  \bibinfo{journal}{Phys. Rev. D} \textbf{\bibinfo{volume}{111}},
  \bibinfo{pages}{123006} (\bibinfo{year}{2025}), \eprint{2310.01483}.

\bibitem[{\citenamefont{Essig et~al.}(2016)\citenamefont{Essig,
  Fernandez-Serra, Mardon, Soto, Volansky, and Yu}}]{Essig:2015cda}
\bibinfo{author}{\bibfnamefont{R.}~\bibnamefont{Essig}},
  \bibinfo{author}{\bibfnamefont{M.}~\bibnamefont{Fernandez-Serra}},
  \bibinfo{author}{\bibfnamefont{J.}~\bibnamefont{Mardon}},
  \bibinfo{author}{\bibfnamefont{A.}~\bibnamefont{Soto}},
  \bibinfo{author}{\bibfnamefont{T.}~\bibnamefont{Volansky}}, \bibnamefont{and}
  \bibinfo{author}{\bibfnamefont{T.-T.} \bibnamefont{Yu}},
  \bibinfo{journal}{JHEP} \textbf{\bibinfo{volume}{05}}, \bibinfo{pages}{046}
  (\bibinfo{year}{2016}), \eprint{1509.01598}.

\bibitem[{\citenamefont{Trickle et~al.}(2020)\citenamefont{Trickle, Zhang,
  Zurek, Inzani, and Griffin}}]{Trickle:2019nya}
\bibinfo{author}{\bibfnamefont{T.}~\bibnamefont{Trickle}},
  \bibinfo{author}{\bibfnamefont{Z.}~\bibnamefont{Zhang}},
  \bibinfo{author}{\bibfnamefont{K.~M.} \bibnamefont{Zurek}},
  \bibinfo{author}{\bibfnamefont{K.}~\bibnamefont{Inzani}}, \bibnamefont{and}
  \bibinfo{author}{\bibfnamefont{S.~M.} \bibnamefont{Griffin}},
  \bibinfo{journal}{JHEP} \textbf{\bibinfo{volume}{03}}, \bibinfo{pages}{036}
  (\bibinfo{year}{2020}), \eprint{1910.08092}.

\bibitem[{\citenamefont{Hochberg et~al.}(2021)\citenamefont{Hochberg, Kahn,
  Kurinsky, Lehmann, Yu, and Berggren}}]{Hochberg:2021pkt}
\bibinfo{author}{\bibfnamefont{Y.}~\bibnamefont{Hochberg}},
  \bibinfo{author}{\bibfnamefont{Y.}~\bibnamefont{Kahn}},
  \bibinfo{author}{\bibfnamefont{N.}~\bibnamefont{Kurinsky}},
  \bibinfo{author}{\bibfnamefont{B.~V.} \bibnamefont{Lehmann}},
  \bibinfo{author}{\bibfnamefont{T.~C.} \bibnamefont{Yu}}, \bibnamefont{and}
  \bibinfo{author}{\bibfnamefont{K.~K.} \bibnamefont{Berggren}},
  \bibinfo{journal}{Phys. Rev. Lett.} \textbf{\bibinfo{volume}{127}},
  \bibinfo{pages}{151802} (\bibinfo{year}{2021}), \eprint{2101.08263}.

\bibitem[{\citenamefont{Essig et~al.}(2012)\citenamefont{Essig, Mardon, and
  Volansky}}]{Essig:2011nj}
\bibinfo{author}{\bibfnamefont{R.}~\bibnamefont{Essig}},
  \bibinfo{author}{\bibfnamefont{J.}~\bibnamefont{Mardon}}, \bibnamefont{and}
  \bibinfo{author}{\bibfnamefont{T.}~\bibnamefont{Volansky}},
  \bibinfo{journal}{Phys. Rev. D} \textbf{\bibinfo{volume}{85}},
  \bibinfo{pages}{076007} (\bibinfo{year}{2012}), \eprint{1108.5383}.

\bibitem[{\citenamefont{Wigner}(1959)}]{wignerD}
\bibinfo{author}{\bibfnamefont{E.~P.} \bibnamefont{Wigner}},
  \emph{\bibinfo{title}{Group theory and its application to the quantum
  mechanics of atomic spectra}} (\bibinfo{year}{1959}).

\bibitem[{\citenamefont{Kievsky et~al.}(2008)\citenamefont{Kievsky, Rosati,
  Viviani, Marcucci, and Girlanda}}]{Kievsky:2008es}
\bibinfo{author}{\bibfnamefont{A.}~\bibnamefont{Kievsky}},
  \bibinfo{author}{\bibfnamefont{S.}~\bibnamefont{Rosati}},
  \bibinfo{author}{\bibfnamefont{M.}~\bibnamefont{Viviani}},
  \bibinfo{author}{\bibfnamefont{L.~E.} \bibnamefont{Marcucci}},
  \bibnamefont{and} \bibinfo{author}{\bibfnamefont{L.}~\bibnamefont{Girlanda}},
  \bibinfo{journal}{J. Phys. G} \textbf{\bibinfo{volume}{35}},
  \bibinfo{pages}{063101} (\bibinfo{year}{2008}), \eprint{0805.4688}.

\bibitem[{\citenamefont{Lillard}(2023)}]{Lillard_VSDM_2023}
\bibinfo{author}{\bibfnamefont{B.}~\bibnamefont{Lillard}},
  \emph{\bibinfo{title}{{VSDM: Vector Spaces for Dark Matter}}}
  (\bibinfo{year}{2023}), \urlprefix\url{{https://github.com/blillard/vsdm}}.

\bibitem[{\citenamefont{Fox et~al.}(2011{\natexlab{a}})\citenamefont{Fox,
  Kribs, and Tait}}]{Fox:2010bu}
\bibinfo{author}{\bibfnamefont{P.~J.} \bibnamefont{Fox}},
  \bibinfo{author}{\bibfnamefont{G.~D.} \bibnamefont{Kribs}}, \bibnamefont{and}
  \bibinfo{author}{\bibfnamefont{T.~M.~P.} \bibnamefont{Tait}},
  \bibinfo{journal}{Phys. Rev. D} \textbf{\bibinfo{volume}{83}},
  \bibinfo{pages}{034007} (\bibinfo{year}{2011}{\natexlab{a}}),
  \eprint{1011.1910}.

\bibitem[{\citenamefont{Fox et~al.}(2011{\natexlab{b}})\citenamefont{Fox, Liu,
  and Weiner}}]{Fox:2010bz}
\bibinfo{author}{\bibfnamefont{P.~J.} \bibnamefont{Fox}},
  \bibinfo{author}{\bibfnamefont{J.}~\bibnamefont{Liu}}, \bibnamefont{and}
  \bibinfo{author}{\bibfnamefont{N.}~\bibnamefont{Weiner}},
  \bibinfo{journal}{Phys. Rev. D} \textbf{\bibinfo{volume}{83}},
  \bibinfo{pages}{103514} (\bibinfo{year}{2011}{\natexlab{b}}),
  \eprint{1011.1915}.

\bibitem[{\citenamefont{Gondolo and Gelmini}(2012)}]{Gondolo:2012rs}
\bibinfo{author}{\bibfnamefont{P.}~\bibnamefont{Gondolo}} \bibnamefont{and}
  \bibinfo{author}{\bibfnamefont{G.~B.} \bibnamefont{Gelmini}},
  \bibinfo{journal}{JCAP} \textbf{\bibinfo{volume}{12}}, \bibinfo{pages}{015}
  (\bibinfo{year}{2012}), \eprint{1202.6359}.

\bibitem[{\citenamefont{Kavanagh}(2014)}]{Kavanagh:2013eya}
\bibinfo{author}{\bibfnamefont{B.~J.} \bibnamefont{Kavanagh}},
  \bibinfo{journal}{Phys. Rev. D} \textbf{\bibinfo{volume}{89}},
  \bibinfo{pages}{085026} (\bibinfo{year}{2014}), \eprint{1312.1852}.

\bibitem[{\citenamefont{Feldstein and Kahlhoefer}(2014)}]{Feldstein:2014gza}
\bibinfo{author}{\bibfnamefont{B.}~\bibnamefont{Feldstein}} \bibnamefont{and}
  \bibinfo{author}{\bibfnamefont{F.}~\bibnamefont{Kahlhoefer}},
  \bibinfo{journal}{JCAP} \textbf{\bibinfo{volume}{08}}, \bibinfo{pages}{065}
  (\bibinfo{year}{2014}), \eprint{1403.4606}.

\bibitem[{\citenamefont{Chen et~al.}(2021)\citenamefont{Chen, Gelmini, and
  Takhistov}}]{Chen:2021qao}
\bibinfo{author}{\bibfnamefont{M.}~\bibnamefont{Chen}},
  \bibinfo{author}{\bibfnamefont{G.~B.} \bibnamefont{Gelmini}},
  \bibnamefont{and}
  \bibinfo{author}{\bibfnamefont{V.}~\bibnamefont{Takhistov}},
  \bibinfo{journal}{JCAP} \textbf{\bibinfo{volume}{12}}, \bibinfo{pages}{048}
  (\bibinfo{year}{2021}), \eprint{2105.08101}.

\bibitem[{\citenamefont{Chen et~al.}(2023)\citenamefont{Chen, Gelmini, and
  Takhistov}}]{Chen:2022xzi}
\bibinfo{author}{\bibfnamefont{M.}~\bibnamefont{Chen}},
  \bibinfo{author}{\bibfnamefont{G.~B.} \bibnamefont{Gelmini}},
  \bibnamefont{and}
  \bibinfo{author}{\bibfnamefont{V.}~\bibnamefont{Takhistov}},
  \bibinfo{journal}{Phys. Lett. B} \textbf{\bibinfo{volume}{841}},
  \bibinfo{pages}{137922} (\bibinfo{year}{2023}), \eprint{2209.10902}.

\bibitem[{\citenamefont{Peter}(2011)}]{Peter:2011eu}
\bibinfo{author}{\bibfnamefont{A.~H.~G.} \bibnamefont{Peter}},
  \bibinfo{journal}{Phys. Rev. D} \textbf{\bibinfo{volume}{83}},
  \bibinfo{pages}{125029} (\bibinfo{year}{2011}), \eprint{1103.5145}.

\bibitem[{\citenamefont{Peter et~al.}(2014)\citenamefont{Peter, Gluscevic,
  Green, Kavanagh, and Lee}}]{Peter:2013aha}
\bibinfo{author}{\bibfnamefont{A.~H.~G.} \bibnamefont{Peter}},
  \bibinfo{author}{\bibfnamefont{V.}~\bibnamefont{Gluscevic}},
  \bibinfo{author}{\bibfnamefont{A.~M.} \bibnamefont{Green}},
  \bibinfo{author}{\bibfnamefont{B.~J.} \bibnamefont{Kavanagh}},
  \bibnamefont{and} \bibinfo{author}{\bibfnamefont{S.~K.} \bibnamefont{Lee}},
  \bibinfo{journal}{Phys. Dark Univ.} \textbf{\bibinfo{volume}{5-6}},
  \bibinfo{pages}{45} (\bibinfo{year}{2014}), \eprint{1310.7039}.

\bibitem[{\citenamefont{Lee}(2014)}]{Lee:2014cpa}
\bibinfo{author}{\bibfnamefont{S.~K.} \bibnamefont{Lee}},
  \bibinfo{journal}{JCAP} \textbf{\bibinfo{volume}{03}}, \bibinfo{pages}{047}
  (\bibinfo{year}{2014}), \eprint{1401.6179}.

\end{thebibliography}

\end{document}